\documentclass[aps,pra,amsmath,onecolumn,amssymb,showpacs]{revtex4}
\usepackage{graphicx,epsfig,epsf}
\usepackage{dcolumn}
\usepackage{bm}
\usepackage{hyperref}

\begin{document}
\newcommand{\ri}{{\rm i}}
\newcommand{\re}{{\rm e}}
\newcommand{\bb}{{\bf b}}
\newcommand{\bc}{{\bf c}}
\newcommand{\bx}{{\bf x}}
\newcommand{\bz}{{\bf z}}
\newcommand{\by}{{\bf y}}
\newcommand{\bv}{{\bf v}}
\newcommand{\bd}{{\bf d}}
\newcommand{\br}{{\bf r}}
\newcommand{\bk}{{\bf k}}
\newcommand{\bA}{{\bf A}}
\newcommand{\bE}{{\bf E}}
\newcommand{\bF}{{\bf F}}
\newcommand{\bI}{{\bf I}}
\newcommand{\bR}{{\bf R}}
\newcommand{\bM}{{\bf M}}
\newcommand{\bX}{{\bf X}}
\newcommand{\bn}{{\bf n}}
\newcommand{\bs}{{\bf s}}
\newcommand{\tr}{{\rm tr}}
\newcommand{\tbs}{\tilde{\bf s}}
\newcommand{\rSi}{{\rm Si}}
\newcommand{\beps}{\mbox{\boldmath{$\epsilon$}}}
\newcommand{\bthe}{\mbox{\boldmath{$\theta$}}}
\newcommand{\blam}{\mbox{\boldmath{$\lambda$}}}
\newcommand{\rg}{{\rm g}}
\newcommand{\xmax}{x_{\rm max}}
\newcommand{\ra}{{\rm a}}
\newcommand{\rx}{{\rm x}}
\newcommand{\rs}{{\rm s}}
\newcommand{\rP}{{\rm P}}
\newcommand{\up}{\uparrow}
\newcommand{\down}{\downarrow}
\newcommand{\hc}{H_{\rm cond}}
\newcommand{\kb}{k_{\rm B}}
\newcommand{\cI}{{\cal I}}
\newcommand{\tit}{\tilde{t}}
\newcommand{\cE}{{\cal E}}
\newcommand{\cC}{{\cal C}}
\newcommand{\Ubs}{U_{\rm BS}}
\newcommand{\qq}{{\bf ???}}
\newcommand*{\etal}{\textit{et al.}}
\newcommand{\gamf}{}
\newcommand{\gamfb}{}

\newcommand{\comment}[1]{{\bf #1}}

\def\vec#1{\mathbf{#1}}
\def\ket#1{|#1\rangle}
\def\bra#1{\langle#1|}
\def\ketbra#1{|#1\rangle\langle#1|}
\newcommand{\braket}[2]{\langle#1|#2\rangle}
\newcommand{\scalp}[2]{\langle#1|#2\rangle}
\def\idmat{\mathbf{1}}
\def\caln{\mathcal{N}}
\def\calc{\mathcal{C}}
\def\rhon{\rho_{\mathcal{N}}}
\def\rhoc{\rho_{\mathcal{C}}}
\def\tr{\mathrm{tr}}
\def\bfu{\mathbf{u}}
\def\bfmu{\mbox{\boldmath$\mu$}}

\newcommand{\be}{\begin{equation}}
\newcommand{\ee}{\end{equation}}
\newcommand{\bfg}{\begin{figure}}
\newcommand{\efg}{\end{figure}}
\newcommand{\Itwo}{\mathbb{1}_2}
\newcommand{\I}{\mathcal{I}}
\newcommand{\al}{\alpha}
\bibliographystyle{apsrev}

\sloppy

\title{Quantum Data-Fitting}
\author{ Nathan Wiebe$^1$, Daniel Braun$^{2,3}$, and Seth Lloyd$^4$ 
}
\affiliation{$^1$ Institute for Quantum Computing, Waterloo, {\cal O}n, Canada}
\affiliation{$^2$ Universit\'e de Toulouse, UPS, Laboratoire
de Physique Th\'eorique (IRSAMC), F-31062 Toulouse, France}
\affiliation{$^3$ CNRS, LPT (IRSAMC), F-31062 Toulouse, France}\
\affiliation{$^4$ MIT - Research Laboratory for Electronics and
  Department of Mechanical Engineering, Cambridge, MA 02139, USA}

\begin{abstract}
We provide a new quantum algorithm that efficiently determines the quality
of a least-squares fit over an  
exponentially large data set by building upon an
algorithm for solving systems of linear 
equations efficiently (Harrow et al., 
Phys.~Rev.~Lett.~{\bf 103}, 150502 (2009)).  In many
cases, our algorithm can also efficiently find a concise function that
approximates the data to be fitted
and bound the approximation error. In cases where the input data is a pure quantum state,
the algorithm can be used to provide an efficient parametric estimation of the quantum state and therefore can be
applied as an alternative to full quantum state tomography given a fault tolerant quantum computer.
\end{abstract}
\maketitle

Invented as early as 1794 by Carl Friedrich Gauss, fitting data to
theoretical models has become over the centuries one of the most important
tools in all  
of quantitative science \cite{Bretscher95}.  Typically, a theoretical model
depends on a number 
of parameters, and leads to functional relations between data that will depend
on those parameters.  Fitting a large amount of experimental data to the
functional 
relations allows one to obtain reliable estimates of the parameters.  If the
amount of data becomes very large, fitting can become very costly.  Examples
include inversion problems of X-ray or
neutron scattering data for structure analysis, or high-energy physics with
giga-bytes of data 
produced per second at the LHC. Typically, structure analysis starts from a
first guess of the structure, and then iteratively tries to improve the
fit to the experimental data by testing variations of the structure. It is
therefore often desirable to test many {\em different 
models}, and compare the best possible fits they provide before committing
to one for which one extracts then the parameters from the fit. Obtaining a
good fit with a relatively small number of parameters compared to the amount
of data can be considered a form of data compression.  Indeed, also for
numerically calculated data, such as many-body wave-functions in molecular
engineering,  efficient fitting of the wave-functions to simpler models would
be highly desirable.  

With the rise of quantum information theory, one might
wonder if a quantum algorithm can be found that solves these problems
efficiently. The discovery that exploiting 
quantum mechanical effects  might lead to enhanced computational power
compared to 
classical information processing has triggered large-scale research
aimed at finding quantum algorithms which are more efficient
than the best classical counterparts
\cite{Shor94,Simon94,Grover97,Dam01,Aharonov06,Childs02}.  Although
fault--tolerant quantum computation remains out of reach at present,
quantum simulation is already now on the verge of providing answers to
questions concerning the states of complex systems that are beyond classical
computability \cite{BMS+11,SBM+11}.
Recently, a quantum algorithm (called HHL in the following) was
introduced that efficiently 
solves a linear equation, $\bF\bx=\bb$, with given vector $\bb$ of dimension
$N$ 
and sparse Hermitian matrix $\bF$ \cite{Harrow09.2}.  ``Efficient
solution'' means
that the expectation value $\bra{\bx}\bM\ket{\bx}$ of an arbitrary
poly-size Hermitian operator $\bM$ can be found in roughly ${
  {\cal O}}(s^4\kappa^2\log(N)/\epsilon)$ steps~\cite{HHLcorrection}, where
$\kappa$ is the condition 
number of $\bF$, i.e.~the ratio between the largest and smallest
eigenvalue of $\bF$, $s$ denotes the sparsenes (i.e.~the maximum number of
non-zero matrix 
elements of
$\bF$ in any given row or column),  and $\epsilon$ is the maximum allowed distance 
between the $\ket{x}$ found by the computer and the exact solution. 
In contrast, they show that it is unlikely that classical computers can efficiently
solve similar problems because it would imply that quantum computers
are no more powerful than classical computers.

While it has remained unclear so far whether expectation values of the form
$\bra{\bf x}{\bf M}\ket{\bf x}$ provide answers to computationally important
  questions, we provide here an adaption of the algorithm to the problem of
  data fitting that allows one to 
efficiently obtain 
the quality of a fit without having to learn the fit-parameters.  {\cal O}ur
algorithm is particularly useful for fitting data {\em
  efficiently computed} by a quantum computer or quantum simulator,
especially if an evolution
can be efficiently simulated but no known method exists to efficiently
learn the resultant state.  
For example, our algorithm could be used to efficiently find
a concise matrix--product state approximation to
a groundstate yielded by a quantum many--body simulator and assess the
 approximation error. 
More complicated states can be used in the fit if the quantum computer
can efficiently prepare them. Fitting quantum states to a set of
known functions is an interesting alternative to performing
full quantum-state tomography \cite{H97}.




\emph{Least-squares fitting}--
The goal in least--squares fitting is to find a simple continuous function that well approximates a discrete set of $N$ points $\{x_i,y_i\}$.  The function is constrained to be linear in the fit parameters
$\blam\in\mathbb{C}^M$, but it can be non-linear in 
$\bx$.  For simplicity we consider
$x\in \mathbb{C}$, but the generalization to higher dimensional $x$ is
straight-forward. 
{\cal O}ur fit function is then of the form $$f(x,\blam):=\sum_{j=1}^M
f_j(x)\lambda_j$$ where $\lambda_j$ is a component of $\blam$ and 
$f(x,\blam):\mathbb{C}^{M+1}\mapsto\mathbb{C}$.  
The optimal fit parameters can be found by minimizing
\begin{equation} \label{E}
E=\sum_{i=1}^N|f(x_i,\blam)-y_i|^2=|\bF\blam-\by|^2
\end{equation}
over all $\blam$, where we have defined the $N\times M$ matrix $\bF$ through
$\bF_{ij}=f_j(x_i)$, $\bF^t$ is its transpose, and $\by$ denotes the column
vector $(y_1,\ldots,y_N)^t$.  Also, following HHL, we assume without loss of generality that $\frac{1}{\kappa^2}\le\|{\bF}^\dagger\bF\|\le 1$ and $\frac{1}{\kappa^2}\le\|{\bF}\bF^\dagger\|\le 1$~\cite{Harrow09.2}.
 Throughout this Letter we use
$\|\cdot\|$ to denote the spectral norm.

Given that $\bF^\dagger\bF$ is invertible, the fit parameters that give
the least square error are 
 found by applying
the Moore--Penrose pseudoinverse \cite{BiG+74} of $\bF$, $\bF^+$, to $\by$:
\begin{equation}
\blam=\bF^+\by=(\bF^\dagger \bF)^{-1}\bF^\dagger \by.\label{eq:pseudoinverse}
\end{equation}
A proof that~\eqref{eq:pseudoinverse} gives an optimal $\blam$ for a least--square fit
 is given in the appendix.

The algorithm consists of three subroutines: a quantum algorithm for performing the pseudo--inverse,
an algorithm for estimating the fit quality and an algorithm for learning
the fit-parameters $\blam$.\\

\emph{1. Fitting Algorithm}---
{\cal O}ur algorithm uses a quantum computer and oracles that
output quantum states that encode the matrix elements of $\bF$ to approximately prepare $\bF^{+}\by$.  The matrix multiplications,
and inversions, are implemented using an improved version of the HHL
algorithm~\cite{Harrow09.2} that utilizes recent developments in
quantum simulation algorithms. 

{\flushleft\emph{Input}: A quantum state 
$\ket{\by}=\sum_{p=M+1}^{M+N} \by_p \ket{p}/|\by|$
that
stores the data $\by$, an upper bound (denoted $\kappa$)
for the square roots of the condition numbers of $\bF\bF^\dagger$ and $\bF^\dagger\bF$,  
the
sparseness  of $\bF$ (denoted $s$) and an error tolerance 
$\epsilon$.}

{\flushleft\emph{Output}: A quantum state $\ket{\blam}$ that is approximately proportional to the
optimal fit parameters  
$\blam/|\blam|$ up to error $\epsilon$ as measured by the Euclidean--norm.}

{\flushleft\emph{Computational Model}: We have a universal quantum computer  equipped
with oracles that, when queried about a non--zero matrix element in a given row, yield a quantum state 
that encodes a requested bit of a binary encoding 
the column number or value of a nonzero matrix element of $\bF$  in a manner 
similar to those in~\cite{WBHS11}. We also assume 
a quantum blackbox  is provided
that yields copies of the input state $\ket{\by}$ on demand.}

{\flushleft \emph{Query Complexity}:  The number of oracle queries used is }
\begin{equation}
\tilde {\cal O}\left(\log(N)(s^3\kappa^6)/\epsilon\right),\label{eq:alg1cost}
\end{equation}
where $\tilde {\cal O}$ notation implies an upper bound on the scaling of a function, suppressing all sub-polynomial functions.
Alternatively, the simulation method
of~\cite{Chi09,BC12} can be used to achieve a query complexity of 
\begin{equation}
\tilde {\cal O}\left( \log(N)(s\kappa^6)/\epsilon^2\right).
\end{equation}

\emph{Analysis of Algorithm}---
The operators $\bF$ and $\bF^{\dagger}$ are implemented using an isometry superoperator $\bI$ to represent them as Hermitian operators on $\mathbb{C}^{N+M}$.
The isometry has the following action on a matrix $\bX$:
\begin{equation}
\bI:\bX\mapsto\left(\begin{array}{cc} 0 &\bX\\ \bX^\dagger & 0 \end{array} \right).\label{iso}
\end{equation}
These choices are convenient because $\bI(\bF^{\dagger})\ket{\by}$ contains $\bF^{\dagger} \by/|\by|$ in its first $M$ entries.
We also assume for simplicity that $|\bI(\bF^\dagger)\ket{\by}|=1$.  This can easily be relaxed by dividing $\bI(\bF^\dagger)\ket{\by}$ by $|\bF^\dagger\by|$.

\emph{Preparing $\bI(\bF^{\dagger})\ket{\by}$}---
The next step is to prepare the state $\bI(\bF^{\dagger})\ket{\by}$.  This is not straightforward because $\bI(\bF^{\dagger})$ is a Hermitian, rather than unitary, operator.
We implement the Hermitian operator using the same phase estimation trick that HHL use to enact the inverse of a Hermitian operator, but instead of
dividing by the eigenvalues of each eigenstate we multiply each eigenstate by its eigenvalue.  We describe the relevant steps below.  For more details, see~\cite{Harrow09.2}.

The algorithm first prepares an ancilla state for a large integer $T$
that is of order $N$
\begin{equation}
\ket{\Psi_0}=\sqrt{\frac{2}{T}}\sum_{\tau=0}^{T-1} \sin \left(\frac{\pi(\tau+1/2)}{T} \right)\ket{\tau}\otimes\ket{\by}.
\end{equation}
It then maps $\ket{\Psi_0}$ to,
\begin{align}
\sqrt{\frac{2}{T}}\sum_{\tau=0}^{T-1} \sin \left(\frac{\pi(\tau+1/2)}{T}\right)\ket{\tau}  \otimes e^{-i\bI(\bF^{\dagger})\tau t_0/T}\ket{\by},\label{eq:phasest1}
\end{align}
for $t_0\in {\cal O}(\kappa/\epsilon)$.  We know from work on quantum simulation
that $\exp(-i\bI(\bF^{\dagger})\tau t_0/T)$ can be implemented within error ${\cal O}(\epsilon)$
in the 2-norm using $\tilde {\cal O}(\log(N)s^3t_0/T)$ quantum operations, if
$\bF$ has sparseness $s$~\cite{CK11}.  Alternatively, the method
of~\cite{Chi09,BC12} gives  
query complexity $\tilde {\cal O}(\log(N)s \tau t_0/(\epsilon T))$.
If we write $\ket{\by}=\sum_{j=1}^N \beta_j \ket{\mu_j}$, where $\ket{\mu_j}$ are the eigenvectors of
$\bI(\bF^\dagger)$ with eigenvalue $E_j$ we obtain
\begin{align}
\sqrt{\frac{2}{T}}\sum_{\tau=0}^{T-1} \sin \left(\frac{\pi(\tau+1/2)}{T}\right)e^{-iE_j\tau t_0/T}\ket{\tau}  \otimes \beta_j\ket{\mu_j},\label{eq:phasest1}
\end{align}
The quantum Fourier transform is then applied to the first register and, after labeling the Fourier coefficients $\alpha_{k|j}$,
the state becomes
\begin{equation}
\sum_{j=1}^N \sum_{k=0}^{T-1} \alpha_{k|j} \beta_j \ket{k}\ket{\mu_j},\label{eq:fouriertrans}
\end{equation}
HHL show that the Fourier
coefficients are small unless the eigenvalue $E_j\approx \tilde E_k:=2\pi
k/t_0$, and $t_0\in {\cal O}(\kappa/\epsilon)$ is needed to ensure that the error from approximating the eigenvalue is at most $\epsilon$.
 It can be seen using the analysis in~\cite{Harrow09.2} that after
re-labeling $\ket{k}$ as $\ket{\tilde E_k}$, and taking $T\in {\cal O}(N)$, \eqref{eq:fouriertrans} is
exponentially close to
$\sum_{j=1}^N \beta_j \ket{\tilde E_j}\ket{\mu_j}$.

The final step is to introduce an ancilla system and perform a controlled
unitary on it that rotates the ancilla state from $\ket{0}$ to
$\sqrt{1-C^2\tilde E_j^2}\ket{0}+C\tilde E_j\ket{1}$, where $C\in {\cal O}(\max_j |E_j|)^{-1}$ because the state would not be properly normalized if $C$ were larger.  The probability of
measuring the ancilla to be $1$ is ${\cal O}(1/\kappa^2)$ since $CE_j$ is at least  ${\cal O}(1/\kappa)$.  ${\cal O}(\kappa^2)$ repetitions are therefore needed to guarantee success with high probability, and amplitude amplification can be used to reduce the number of repetitions to ${\cal O}(\kappa)$~\cite{Harrow09.2}.  HHL show that either ${\cal O}(1/\kappa^2)$ or ${\cal O}(1/\kappa)$ attempts are also needed to successfully perform $\bI(\bF)^{-1}$ depending on whether amplitude amplification is used.

The cost of implementing $\bI(\bF^\dagger)$ is the product of
the cost of simulating $\bI(\bF^\dagger)$ for time $\kappa/\epsilon$ and the number of repetitions required to obtain a successful result, which scales as $O(\kappa)$.
The 
improved simulation 
method of Childs and Kothari~\cite{CK11} allows the simulation to be performed in time $\tilde{\cal O}(\log(N)s^3\kappa/\epsilon)$, where $s$ is the sparseness of $\bF$; therefore, $\bI(\bF^\dagger)\ket{\by}$ can be prepared using $\tilde {\cal O}( \log(N) s^3 \kappa^2/\epsilon)$
oracle calls. 
The cost of performing the inversion using the simulation method of~\cite{Chi09,BC12} is found by substituting $s\rightarrow s^{1/3}/\epsilon$ into this or any of our subsequent results. 

\emph{Inverting $\bF^{\dagger}\bF$}---
We then finish the algorithm by applying $(\bF^{\dagger}\bF)^{-1}$ using the method
of HHL~\cite{Harrow09.2}. Note that the existence of $(\bF^{\dagger}\bF)^{-1}$ is
implied by a well-defined fitting-problem, in the sense that a zero eigenvalue 
of $\bF^{\dagger}\bF$ would result in a degenerate direction of the quadratic form
(\ref{E}). 
The operator $\bF^{\dagger} \bF\in \mathbb{C}^{M\times M}$ is Hermitian and hence amenable to the linear systems algorithm.
We do, however, need to extend the domain of the operator to make it
compatible with $\ket{\by}$ which is in a Hilbert space of dimension $N+M$. 
We introduce $\bA$ to denote the corresponding operator,
\begin{equation}
\bA:= \left(\begin{array}{cc} \bF^{\dagger}\bF &0\\ 0 & \bF\bF^\dagger \end{array} \right)=\bI(\bF)^2.
\end{equation}
If we define $\ket{\blam}\in \mathbb{C}^{N+M}$ to be a state of the form $\ket{\blam}=\sum_{j=1}^{M} \lambda_j\ket{j}$ up to a normalizing constant, then
$\bF^{\dagger} \bF \blam$ is proportional to $\bA \ket{\blam}$ up to a normalizing
constant.  This means that we can find a vector that is proportional 
to the least-squares fit parameters by inversion via
\begin{equation}
 \ket{\blam}= \bA^{-1}\bI(\bF^{\dagger})\ket{\by}.\label{lamend}
\end{equation}
This can be further simplified by noting that
\begin{equation}
\bA^{-1}=\bI(\bF)^{-2}.
\end{equation}
Amplitude amplification does not decrease the number of attempts needed to implement $\bA^{-1}$ in~\eqref{lamend} because the algorithm require reflections about $\bI(\bF^\dagger)\ket{\by}$, which requires ${\cal O}(\kappa)$ repetitions to prepare. 

Since amplitude amplification provides no benefit for implementing $\bA^{-1}$, ${\cal O}(\kappa^5)$ repetitions are needed to implement $\bA^{-1}\bI(\bF^\dagger)$.  This is a consequence of the fact that the probability of successfully performing each $\bI(\bF)^{-1}$ is ${\cal O}(1/\kappa^2)$ and the probability of performing  $\bI(\bF^\dagger)$ is ${\cal O}(1/\kappa)$ (if amplitude amplification is used).  The cost of performing the simulations involved in each attempt is $\tilde{\cal O}(\log(N) s^3\kappa/\epsilon)$ and hence the required number of oracle calls scales as
\begin{equation}
\tilde{{\cal O}}\left(\log(N)(s^3\kappa^6/\epsilon) \right).~\label{eq:totalcost}
\end{equation}

Although the algorithm yields $\ket{\blam}$ efficiently, it may be exponentially expensive to learn
$\ket{\blam}$ via tomography; however, we show below that a quantum computer can assess the quality of the fit efficiently.

\emph{2. Estimating Fit Quality}---
We will now show that we can efficiently estimate
the fit quality $E$ 
even if $M$ is exponentially large and without having to determine the
fit-parameters.  For this problem, note that due to the isometry (\ref{iso})
${E}=|\ket{\by}- 
\bI(\bF)\ket{\blam}|^2$.   We assume the prior computational model. We are
also provided a desired error tolerance, $\epsilon$,  
and wish to determine the quality of the fit within error $\delta$.

{\flushleft\emph{Input}: A constant $\delta>0$ and all inputs required by algorithm 1.}

{\flushleft\emph{Output}: An estimate of $\gamfb|\bra{\by}\bI(\bF)\ket{\blam}|^2$ accurate within error $\delta$.}

{\flushleft\emph{Query Complexity}:
\begin{equation}
\tilde {\cal O}\left(\log(N)\frac{ s^3 \kappa^4}{\epsilon\delta^2}\right).
\end{equation}
}

\emph{Algorithm}---  We begin by preparing the state $\ket{\by}\otimes
\ket{\by}$ using the provided state preparation blackbox.  We then use the prior
algorithm to construct 
the state
\begin{equation}
\bI(\bF)\bA^{-1}\bI(\bF^{\dagger})\ket{\by}\otimes \ket{\by}=\bI(\bF)^{-1}\bI(\bF^{\dagger})\ket{\by}\otimes \ket{\by},\label{eq:deltalearna}
\end{equation}
within error ${\cal O}(\epsilon)$. The cost of implementing $\bI(\bF)^{-1}\bI(\bF^{\dagger})$ (with high probability) within error $\epsilon$ is
\begin{equation}
\tilde {\cal O}\left(\log(N)\frac{ s^3\kappa^4}{\epsilon}\right).\label{eq:deltalearn}
\end{equation}

The swap test~\cite{BCWdW01} is then used to determine the accuracy of
the fit.  The swap test is a method that can be used to distinguish $\ket{\by}$ and $\gamf\bI(\bF)\ket{\blam}$ by performing a swap operation on the
two quantum states controlled by a qubit in the state $(\ket{0}+\ket{1})/\sqrt{2}$.  The Hadamard operation is then applied to the
control qubit and the control qubit is then measured in the computational basis.  The test concludes that the states are different if the outcome
is ``1''.  The probability of observing an outcome of ``1'' is $(1-\gamfb|\bra{\by}\bI(\bF)\ket{\blam}|^2)/2$ for our problem.  

The overlap between the two quantum states can be learned by statistically sampling the outcomes from many instances of the swap
test.  The value of $\gamfb|\bra{\by}\bI(\bF)\ket{\blam}|^2$ can be approximated using the sample mean of this distribution.  It follows from estimates of the standard deviation of the mean that ${\cal O}(1/\delta^2)$ samples are required to estimate the mean within error
${\cal O}(\delta)$.  The cost of algorithm 2 is then found by multiplying~\eqref{eq:deltalearn} by $1/\delta^2$.

The quantity $E$ can be estimated from the output of algorithm 2 by $E\le 2(1-\gamf|\bra{\by}\bI(\bF)\ket{\blam}|)$.  Taylor series analysis shows
that the error in the upper bound for $E$ is also ${\cal O}(\delta)$.

There are several important limitations to this technique.  First, if $\bF$ is not sparse (meaning $s\in {\cal O}({\rm poly}(N))$) then the algorithm may not be efficient
because the quantum simulation step used in the algorithm may not be efficient.
As noted in previous results~\cite{Chi09,BC12,WBHS11}, we can generalize our results to systems where $\bF$ is non-sparse if there exists a set of efficient unitary transformations
$U_j$ such that $\bI(\bF)=\sum_j U_j H_j U_j^\dagger$ where each $H_j$ is sparse and Hermitian.  Also, in many important cases (such as fitting to experimental data) it may not be posible to prepare the initial 
state $\ket{\by}$ efficiently.  For this reason, our algorithm is better suited for approximating
the output of quantum devices than the classical outputs of
experiments.  Finally, algorithm 2 only provides an efficient 
estimate of the fit quality and does not provide $\blam$;
however, we can use it to determine whether 
a quantum state has a concise representation within a family of states.  If algorithm 2 can be used to find such a representation, then the parameters
$\ket{\blam}$ can be learned via state tomography.  We discuss this
approach below.

{\emph{3. Learning $\blam$}-- This method can also be used to find a concise fit function that approximates $\by$.  Specifically, we use statistical sampling
and quantum state tomography to find a concise representation for the quantum state using $M'$ parameters.  The resulting
algorithm is efficient if $M'\in {\cal O}({\rm polylog}(N))$.}

{\flushleft\emph{Input}: As algorithm 2, but in addition with an integer $M'\in {\cal O}({\rm polylog}(M))$ that gives the
maximum number of fit functions allowed in the fit.}

{\flushleft\emph{Output}: A classical bit string approximating $\ket{\blam}$ to precision $\epsilon$, a list of the $M'$ fit functions that comprise $\ket{\blam}$ and $|\bra{\by}\bI(\bF)\ket{\blam}|^2$ calculated to precision $\delta$.}

{\flushleft\emph{Computational Model}: As algorithm 1, but the oracles can be controlled to either fit the state to all $M$ fit functions or any
subset consisting of $M'$ fit functions.}

{\flushleft \emph{Query Complexity}: $$\tilde {\cal O}\left( \log(N)s^3\left(\frac{\kappa^4}{\epsilon\delta^2}+\frac{M'^2\kappa^6}{\epsilon^3}\right)\right).$$}

\emph{Algorithm}---
The first step of the algorithm is to prepare the state $\ket{\blam}$ using algorithm 1.  The state is then measured ${\cal O}(M')$ times and a histogram
of the measurement outcomes is constructed.  Since the probability of measuring each of these outcomes is proportional to their relevance to the fit, we are
likely to find the $M'$ of the most likely outcomes by sampling the state ${\cal O}(M')$ times.

After choosing the $M'$ most significant fit functions, we remove all other fit functions from the fit and prepare the state $\ket{\blam}$ using the reduced set of
fit functions.  Compressed sensing~\cite{GYF+10,SKM+11,SML+10} is then used to reconstruct $\ket{\blam}$ within ${\cal O}(\epsilon)$ error.  The idea of compressed sensing is that a low--rank density matrix can be uniquely determined (with high probability) by a small number of randomly chosen measurements.  A convex optimization routine is then used to reconstruct the density matrix from the expectation values found for each of the measurements.

Compressed sensing requires ${\cal O}(M' \log(M')^2)$ measurement
settings to reconstruct pure states, and observation 1 of~\cite{GYF+10} implies that ${\cal O}(M'/\epsilon^2)$ measurements are needed for each setting to ensure that the reconstruction error is ${\cal O}(\epsilon)$;  therefore, ${\cal O}(M'^2 \log(M')^2/\epsilon^2)$ measurements are needed to approximate the state within error ${\cal O}(\epsilon)$.  The total cost of learning $\ket{\blam}$
is the number of measurements needed for tomography multiplied by the cost of preparing the state and thus scales as
\begin{equation}
\tilde {\cal O}\left(\log(N)\frac{ s^3M'^2\kappa^6}{\epsilon^3}\right),\label{eq:delta4}
\end{equation}
which subsumes the cost of measuring $\ket{\blam}$ to find the most significant $M'$ fit functions.

Finally, we measure the quality of the fit using algorithm 2.  The total cost of estimating $\ket{\blam}$ and the fit quality is thus the sum of~\eqref{eq:delta4} and~\eqref{eq:deltalearn}, as claimed.

\emph{Remark}: The quality of the resulting fit that is yielded by this algorithm depends strongly on the set of fit functions that are used.  If the fit functions are
chosen well,  fewer than $M'$ fit functions are used to estimate $\ket{\by}$ with high fidelity.  Conversely,  ${\cal O}(N)$ fit functions may be needed to achieve the desired error tolerance if the fit functions are chosen 
poorly.  
Fortunately, the efficiency of algorithm 2 allows the
user to search many sets of possible fit functions for a concise and accurate model within a large set of potential models.

{\em Acknowledgements: } DB would like to thank the Joint Quantum Institute
(NIST and University of Maryland) and the Institute for Quantum Computing
(University of Waterloo), for
hospitality, and Arram Harrow and Avinatan Hassidim for useful
correspondence.  NW would like to thank Andrew Childs and Dominic Berry for useful
feedback and acknowledges support from USAR{\cal O}/DT{\cal O}.  SL is supported by NSF, DARPA, ENI and ISI.

\appendix
\section{Moore--Penrose Pseudoinverse}
Here we review an elementary proof~\cite{Pen56} of why applying Moore--Penrose pseudoinverse to the complex--valued vector $\by$ yields
parameters that minimize the least--squares fit of the initial state.  To begin, we need to prove some properties of the pseudoinverse.
First,
\begin{equation}
(\bF\bF^+)^\dagger = \bF \bF^+.\label{eq:suppprop1}
\end{equation}
The proof of this property is
\begin{equation}
(\bF\bF^+)^\dagger = (\bF (\bF^\dagger\bF)^{-1}\bF^\dagger)^{\dagger}=(\bF ((\bF^\dagger\bF)^{-1})^\dagger\bF^\dagger).
\end{equation}
The result of~\eqref{eq:suppprop1} then follows by noting that $\bF^\dagger\bF$ is self--adjoint.

Next, we need the property that
\begin{equation}
\bF\bF^+\bF=\bF.\label{eq:supprop2}
\end{equation}
This property follows directly from substituting in the definition of $\bF^+$ into the expression.

The final property we need is
\begin{equation}
\bF^\dagger(\bF\bF^+\by-\by)=0.\label{eq:suppprop3}
\end{equation}
Using property~\eqref{eq:suppprop1} we find that
\begin{equation}
\bF^\dagger(\bF\bF^+\by-\by)=(\bF\bF^+\bF)^\dagger\by - \bF^\dagger\by.
\end{equation}
Property~\eqref{eq:supprop2} then implies that
\begin{equation}
(\bF\bF^+\bF)^\dagger\by - \bF^\dagger\by=\bF^\dagger\by - \bF^\dagger\by=0.
\end{equation}

For simplicity, we will express $\bz=\bF^+\by$ and then find
\begin{equation}
\|\bF \blam - \by\|^2 = \|\bF\bz - \by +(\bF\blam-\bF\bz)\|^2.
\end{equation}
Expanding this relation yields,
\begin{widetext}
\begin{equation}
\|\bF \blam - \by\|^2=\|\bF\bz-\by\|^2 + \|\bF(\blam-\bz)\|^2+(\bF\bz-\by)^\dagger\bF(\blam-\bz)+(\blam-\bz)^\dagger\bF^\dagger(\bF\bz-\by).
\end{equation}
\end{widetext}
Property~\eqref{eq:suppprop3} then implies that $\bF^\dagger(\bF\bz-\by)=0$ and hence
\begin{align}
\|\bF \blam - \by\|^2&=\|\bF\bz-\by\|^2 + \|\bF(\blam-\bz)\|^2\nonumber\\
&\ge \|\bF\bF^+\by-\by\|,
\end{align}
which holds with equality if $\blam=\bz=\bF^+\by$.  Therefore, applying the Moore--Penrose peudoinverse to $\by$ provides 
fit parameters that minimize the least--square error.


\begin{thebibliography}{22}
\expandafter\ifx\csname natexlab\endcsname\relax\def\natexlab#1{#1}\fi
\expandafter\ifx\csname bibnamefont\endcsname\relax
  \def\bibnamefont#1{#1}\fi
\expandafter\ifx\csname bibfnamefont\endcsname\relax
  \def\bibfnamefont#1{#1}\fi
\expandafter\ifx\csname citenamefont\endcsname\relax
  \def\citenamefont#1{#1}\fi
\expandafter\ifx\csname url\endcsname\relax
  \def\url#1{\texttt{#1}}\fi
\expandafter\ifx\csname urlprefix\endcsname\relax\def\urlprefix{URL }\fi
\providecommand{\bibinfo}[2]{#2}
\providecommand{\eprint}[2][]{\url{#2}}

\bibitem[{\citenamefont{Bretscher}(1995)}]{Bretscher95}
\bibinfo{author}{\bibfnamefont{{\cal O}.}~\bibnamefont{Bretscher}},
  \emph{\bibinfo{title}{Linear Algebra With Applications, 3rd ed..}}
  (\bibinfo{publisher}{Prentice Hall, Upper Saddle River NJ},
  \bibinfo{year}{1995}).

\bibitem[{\citenamefont{Shor}(1994)}]{Shor94}
\bibinfo{author}{\bibfnamefont{P.~W.} \bibnamefont{Shor}}, in
  \emph{\bibinfo{booktitle}{Proc. 35th Annu. Symp. Foundations of Computer
  Science (ed. Goldwasser, S.), {\em p.~124-134}}} (\bibinfo{publisher}{IEEE
  Computer Society, Los Alamitos, CA}, \bibinfo{year}{1994}).

\bibitem[{\citenamefont{Simon}(1994)}]{Simon94}
\bibinfo{author}{\bibfnamefont{D.}~\bibnamefont{Simon}}
  (\bibinfo{publisher}{(IEEE Computer Society, Los Alamitos, CA},
  \bibinfo{year}{1994}).

\bibitem[{\citenamefont{Grover}(1997)}]{Grover97}
\bibinfo{author}{\bibfnamefont{L.~K.} \bibnamefont{Grover}},
  \bibinfo{journal}{Phys. Rev. Lett.} \textbf{\bibinfo{volume}{79}},
  \bibinfo{pages}{325} (\bibinfo{year}{1997}).

\bibitem[{\citenamefont{van Dam and Hallgren}()}]{Dam01}
\bibinfo{author}{\bibfnamefont{W.}~\bibnamefont{van Dam}} \bibnamefont{and}
  \bibinfo{author}{\bibfnamefont{S.}~\bibnamefont{Hallgren}},
  \emph{\bibinfo{title}{Efficient quantum algorithms for shifted quadratic
  character problems}}, \eprint{arXiv:quant-ph/0011067v2}.

\bibitem[{\citenamefont{Aharonov et~al.}()\citenamefont{Aharonov, Landau, and
  Makowsky}}]{Aharonov06}
\bibinfo{author}{\bibfnamefont{D.}~\bibnamefont{Aharonov}},
  \bibinfo{author}{\bibfnamefont{Z.}~\bibnamefont{Landau}}, \bibnamefont{and}
  \bibinfo{author}{\bibfnamefont{J.}~\bibnamefont{Makowsky}},
  \eprint{quant-ph/0611156}.

\bibitem[{\citenamefont{Childs et~al.}(2002)\citenamefont{Childs, Cleve,
  Deotto, Farhi, Gutmann, and Spielman}}]{Childs02}
\bibinfo{author}{\bibfnamefont{A.~M.} \bibnamefont{Childs}},
  \bibinfo{author}{\bibfnamefont{R.}~\bibnamefont{Cleve}},
  \bibinfo{author}{\bibfnamefont{E.}~\bibnamefont{Deotto}},
  \bibinfo{author}{\bibfnamefont{E.}~\bibnamefont{Farhi}},
  \bibinfo{author}{\bibfnamefont{S.}~\bibnamefont{Gutmann}}, \bibnamefont{and}
  \bibinfo{author}{\bibfnamefont{D.~A.} \bibnamefont{Spielman}},
  \bibinfo{journal}{Proc. 35th ACM Symposium on Theory of Computing (ST{\cal O}C
  2003)} pp. \bibinfo{pages}{59--68} (\bibinfo{year}{2002}).

\bibitem[{\citenamefont{{Barreiro} et~al.}(2011)\citenamefont{{Barreiro},
  {M{\"u}ller}, {Schindler}, {Nigg}, {Monz}, {Chwalla}, {Hennrich}, {Roos},
  {Zoller}, and {Blatt}}}]{BMS+11}
\bibinfo{author}{\bibfnamefont{J.~T.} \bibnamefont{{Barreiro}}},
  \bibinfo{author}{\bibfnamefont{M.}~\bibnamefont{{M{\"u}ller}}},
  \bibinfo{author}{\bibfnamefont{P.}~\bibnamefont{{Schindler}}},
  \bibinfo{author}{\bibfnamefont{D.}~\bibnamefont{{Nigg}}},
  \bibinfo{author}{\bibfnamefont{T.}~\bibnamefont{{Monz}}},
  \bibinfo{author}{\bibfnamefont{M.}~\bibnamefont{{Chwalla}}},
  \bibinfo{author}{\bibfnamefont{M.}~\bibnamefont{{Hennrich}}},
  \bibinfo{author}{\bibfnamefont{C.~F.} \bibnamefont{{Roos}}},
  \bibinfo{author}{\bibfnamefont{P.}~\bibnamefont{{Zoller}}}, \bibnamefont{and}
  \bibinfo{author}{\bibfnamefont{R.}~\bibnamefont{{Blatt}}},
  \bibinfo{journal}{Nature} \textbf{\bibinfo{volume}{470}},
  \bibinfo{pages}{486} (\bibinfo{year}{2011}).

\bibitem[{\citenamefont{{Simon} et~al.}(2011)\citenamefont{{Simon}, {Bakr},
  {Ma}, {Tai}, {Preiss}, and {Greiner}}}]{SBM+11}
\bibinfo{author}{\bibfnamefont{J.}~\bibnamefont{{Simon}}},
  \bibinfo{author}{\bibfnamefont{W.~S.} \bibnamefont{{Bakr}}},
  \bibinfo{author}{\bibfnamefont{R.}~\bibnamefont{{Ma}}},
  \bibinfo{author}{\bibfnamefont{M.~E.} \bibnamefont{{Tai}}},
  \bibinfo{author}{\bibfnamefont{P.~M.} \bibnamefont{{Preiss}}},
  \bibnamefont{and}
  \bibinfo{author}{\bibfnamefont{M.}~\bibnamefont{{Greiner}}},
  \bibinfo{journal}{Nature} \textbf{\bibinfo{volume}{472}},
  \bibinfo{pages}{307} (\bibinfo{year}{2011}).

\bibitem[{\citenamefont{Harrow et~al.}(2009)\citenamefont{Harrow, Hassidim, and
  Lloyd}}]{Harrow09.2}
\bibinfo{author}{\bibfnamefont{A.~W.} \bibnamefont{Harrow}},
  \bibinfo{author}{\bibfnamefont{A.}~\bibnamefont{Hassidim}}, \bibnamefont{and}
  \bibinfo{author}{\bibfnamefont{S.}~\bibnamefont{Lloyd}},
  \bibinfo{journal}{Phys. Rev. Lett.} \textbf{\bibinfo{volume}{103}},
  \bibinfo{pages}{150502} (\bibinfo{year}{2009}).

\bibitem[{HHL()}]{HHLcorrection}
\bibinfo{note}{The result in \cite{Harrow09.2} incorrectly cites the results of
  Theorem 2 of~\cite{BACS07} leading to the conclusion that the cost of linear
  inversion scales with the sparseness as $\tilde {\cal O}(s^2)$ rather than $\tilde
  {\cal O}(s^4)$.}

\bibitem[{\citenamefont{Hradil}(1997)}]{H97}
\bibinfo{author}{\bibfnamefont{Z.}~\bibnamefont{Hradil}},
  \bibinfo{journal}{Phys. Rev. A} \textbf{\bibinfo{volume}{55}},
  \bibinfo{pages}{R1561} (\bibinfo{year}{1997}).

\bibitem[{\citenamefont{Ben-israel et~al.}(1974)\citenamefont{Ben-israel,
  Greville, Ahlberg, Nilson, and Walsh}}]{BiG+74}
\bibinfo{author}{\bibfnamefont{A.}~\bibnamefont{Ben-israel}},
  \bibinfo{author}{\bibfnamefont{T.~N.~E.} \bibnamefont{Greville}},
  \bibinfo{author}{\bibfnamefont{J.~H.} \bibnamefont{Ahlberg}},
  \bibinfo{author}{\bibfnamefont{E.~N.} \bibnamefont{Nilson}},
  \bibnamefont{and} \bibinfo{author}{\bibfnamefont{J.~L.} \bibnamefont{Walsh}},
  pp. \bibinfo{pages}{104--106} (\bibinfo{year}{1974}).

\bibitem[{\citenamefont{Wiebe et~al.}(2011)\citenamefont{Wiebe, Berry,
  H{\o}yer, and Sanders}}]{WBHS11}
\bibinfo{author}{\bibfnamefont{N.}~\bibnamefont{Wiebe}},
  \bibinfo{author}{\bibfnamefont{D.~W.} \bibnamefont{Berry}},
  \bibinfo{author}{\bibfnamefont{P.}~\bibnamefont{H{\o}yer}}, \bibnamefont{and}
  \bibinfo{author}{\bibfnamefont{B.~C.} \bibnamefont{Sanders}},
  \bibinfo{journal}{J. Phys. A} \textbf{\bibinfo{volume}{\textbf{43}}},
  \bibinfo{pages}{065203} (\bibinfo{year}{2011}).

\bibitem[{\citenamefont{Childs}(2009)}]{Chi09}
\bibinfo{author}{\bibfnamefont{A.~M.} \bibnamefont{Childs}},
  \bibinfo{journal}{Commun. Math. Phys.}
  \textbf{\bibinfo{volume}{\textbf{294}}}, \bibinfo{pages}{581}
  (\bibinfo{year}{2009}).

\bibitem[{\citenamefont{Berry and Childs}(2012)}]{BC12}
\bibinfo{author}{\bibfnamefont{D.~W.} \bibnamefont{Berry}} \bibnamefont{and}
  \bibinfo{author}{\bibfnamefont{A.~M.} \bibnamefont{Childs}},
  \bibinfo{journal}{Quantum Information and Computation}
  \textbf{\bibinfo{volume}{12}}, \bibinfo{pages}{29} (\bibinfo{year}{2012}).

\bibitem[{\citenamefont{Childs and Kothari}(2011)}]{CK11}
\bibinfo{author}{\bibfnamefont{A.}~\bibnamefont{Childs}} \bibnamefont{and}
  \bibinfo{author}{\bibfnamefont{R.}~\bibnamefont{Kothari}}, in
  \emph{\bibinfo{booktitle}{Theory of Quantum Computation, Communication, and
  Cryptography}}, edited by \bibinfo{editor}{\bibfnamefont{W.}~\bibnamefont{van
  Dam}}, \bibinfo{editor}{\bibfnamefont{V.}~\bibnamefont{Kendon}},
  \bibnamefont{and} \bibinfo{editor}{\bibfnamefont{S.}~\bibnamefont{Severini}}
  (\bibinfo{publisher}{Springer Berlin / Heidelberg}, \bibinfo{year}{2011}),
  vol. \bibinfo{volume}{6519} of \emph{\bibinfo{series}{Lecture Notes in
  Computer Science}}, pp. \bibinfo{pages}{94--103}, ISBN
  \bibinfo{isbn}{978-3-642-18072-9}.

\bibitem[{\citenamefont{Buhrman et~al.}(2001)\citenamefont{Buhrman, Cleve,
  Watrous, and de~Wolf}}]{BCWdW01}
\bibinfo{author}{\bibfnamefont{H.}~\bibnamefont{Buhrman}},
  \bibinfo{author}{\bibfnamefont{R.}~\bibnamefont{Cleve}},
  \bibinfo{author}{\bibfnamefont{J.}~\bibnamefont{Watrous}}, \bibnamefont{and}
  \bibinfo{author}{\bibfnamefont{R.}~\bibnamefont{de~Wolf}},
  \bibinfo{journal}{Phys. Rev. Lett.} \textbf{\bibinfo{volume}{87}},
  \bibinfo{pages}{167902} (\bibinfo{year}{2001}).

\bibitem[{\citenamefont{Gross et~al.}(2010)\citenamefont{Gross, Liu, Flammia,
  Becker, and Eisert}}]{GYF+10}
\bibinfo{author}{\bibfnamefont{D.}~\bibnamefont{Gross}},
  \bibinfo{author}{\bibfnamefont{Y.-K.} \bibnamefont{Liu}},
  \bibinfo{author}{\bibfnamefont{S.~T.} \bibnamefont{Flammia}},
  \bibinfo{author}{\bibfnamefont{S.}~\bibnamefont{Becker}}, \bibnamefont{and}
  \bibinfo{author}{\bibfnamefont{J.}~\bibnamefont{Eisert}},
  \bibinfo{journal}{Phys. Rev. Lett.} \textbf{\bibinfo{volume}{105}},
  \bibinfo{pages}{150401} (\bibinfo{year}{2010}).

\bibitem[{\citenamefont{Shabani et~al.}(2011)\citenamefont{Shabani, Kosut,
  Mohseni, Rabitz, Broome, Almeida, Fedrizzi, and White}}]{SKM+11}
\bibinfo{author}{\bibfnamefont{A.}~\bibnamefont{Shabani}},
  \bibinfo{author}{\bibfnamefont{R.~L.} \bibnamefont{Kosut}},
  \bibinfo{author}{\bibfnamefont{M.}~\bibnamefont{Mohseni}},
  \bibinfo{author}{\bibfnamefont{H.}~\bibnamefont{Rabitz}},
  \bibinfo{author}{\bibfnamefont{M.~A.} \bibnamefont{Broome}},
  \bibinfo{author}{\bibfnamefont{M.~P.} \bibnamefont{Almeida}},
  \bibinfo{author}{\bibfnamefont{A.}~\bibnamefont{Fedrizzi}}, \bibnamefont{and}
  \bibinfo{author}{\bibfnamefont{A.~G.} \bibnamefont{White}},
  \bibinfo{journal}{Phys. Rev. Lett.} \textbf{\bibinfo{volume}{106}},
  \bibinfo{pages}{100401} (\bibinfo{year}{2011}).

\bibitem[{\citenamefont{Shabani et~al.}(2010)\citenamefont{Shabani, Mohseni,
  Lloyd, Kosut, and Rabitz}}]{SML+10}
\bibinfo{author}{\bibfnamefont{A.}~\bibnamefont{Shabani}},
  \bibinfo{author}{\bibfnamefont{M.}~\bibnamefont{Mohseni}},
  \bibinfo{author}{\bibfnamefont{S.}~\bibnamefont{Lloyd}},
  \bibinfo{author}{\bibfnamefont{R.~L.} \bibnamefont{Kosut}}, \bibnamefont{and}
  \bibinfo{author}{\bibfnamefont{H.}~\bibnamefont{Rabitz}},
  \bibinfo{journal}{arXiv:1002.1330}  (\bibinfo{year}{2010}).

\bibitem[{\citenamefont{Berry et~al.}(2007)\citenamefont{Berry, Ahokas, Cleve,
  and Sanders}}]{BACS07}
\bibinfo{author}{\bibfnamefont{D.~W.} \bibnamefont{Berry}},
  \bibinfo{author}{\bibfnamefont{G.}~\bibnamefont{Ahokas}},
  \bibinfo{author}{\bibfnamefont{R.}~\bibnamefont{Cleve}}, \bibnamefont{and}
  \bibinfo{author}{\bibfnamefont{B.~C.} \bibnamefont{Sanders}},
  \bibinfo{journal}{Commun. Math. Phys.}
  \textbf{\bibinfo{volume}{\textbf{270}}}, \bibinfo{pages}{359}
  (\bibinfo{year}{2007}).

\bibitem[{\citenamefont{Penrose}(2007)\citenamefont{R. Penrose}}]{Pen56}
\bibinfo{author}{\bibfnamefont{R.} \bibnamefont{Penrose}},
  \bibinfo{journal}{Proceedings of The Cambridge Philosophical Society}
  \textbf{\bibinfo{volume}{\textbf{52}}}, \bibinfo{pages}{17}
  (\bibinfo{year}{1956}).

\end{thebibliography}

\end{document}